# The impact of climate and wealth on energy consumption in small tropical islands

Julien Gargani[1,2]

[1]Université Paris-Saclay, Geops, CNRS, 91405 Orsay, France

[2]Université Paris-Saclay, Centre d'Alembert, 91405 Orsay, France

## Abstract

Anthropic activities have a significant causal effect on climatic change but climate has also major impact on human societies. Population vulnerability to natural hazards and limited natural resources are deemed problematic, particularly on small tropical islands. Lifestyles and activities are heavily reliant on energy consumption. The relationship between climatic variations and energy consumption must be clearly understood. We demonstrate that it is possible to determine the impact of climate change on energy consumption. In small tropical islands, the relationship between climate and energy consumption is primarily driven by air conditioner electricity consumption during hotter months. Temperatures above 26°C correlate with increased electricity consumption. Energy consumption is sensitive to: (1) climatic seasonal fluctuations, (2) cyclonic activity, (3) temperature warming over the last 20 years. On small tropical islands, demographic and wealth variations also have a significant impact on energy consumption. The relationship between climate and energy consumption suggests reconsidering the production and consumption of carbon-based energy.

**Highlights:**

- Climate warming causes energy consumption increase in tropical areas
- Seasonal temperature >26°C causes energy consumption increase
- Air conditioners are responsible of energy consumption growth
- Wealth favor energy consumption increase
- Cyclone occurrence causes energy consumption decrease

## 1. Introduction

Multiple environmental [IPBES, 2019], climatic [IPCC, 2013], social and economic crises [Latouche, 2004; Gargani, 2016a] have been described during the last decades. Understanding these crises and the complex interactions between eco-socio systems is of fundamental interest. The observed climate change is due to anthropogenic factors [IPCC, 2013]. It is associated with temperature increases, sea level rises and more intense extreme events [Coumou et al., 2012]. Greenhouse gases, primarily from the production and use of carbon-based energy sources (coal, gas, and oil), are the primary cause of global warming [IPCC, 2013]. Energy consumption is closely linked to our lifestyles [Jones et al., 2015; Pettifor et al., 2023]. If global warming is a consequence of anthropic activities, global warming could also have increasing consequences on energy consumption and on our way of life [Khan et al., 2016].

Energy production is significant and has increased over the last centuries [Fressoz, 2024]. Many social and economic activities require the use of energy. Energy is used for: (i) production (food, industry, raw material extraction, etc.), (ii) transportation, but also (iii) office activities (light, computers, building heating and air conditioning), and (iv) homework and way of life [Chen and Chen, 2011; Syvitsky et al., 2020].

According to Hall and Klitgaard (2011), one indicator that may be used to describe the evolution of human activities is the amount of energy produced and consumed. For example, the Human Development Index (HDI) and the Night Light Devepment Index (NLDI) [Elvidge et al., 2012] are linked. Energy consumption variations has been monitored for decades and observations show that variations depend on: (i) the time of the day, (ii) the day of the week, specifically between workday and weekend, (iii) the season [Li et al., 2018]. Energy consumption is also known to have evolved significantly during the covid-19 pandemic crisis in 2020 [Garcia et al., 2021; Halbrügge et al., 2021; Bertram et al., 2021; Jiang et al., 2021; Navon et al., 2021: Gargani, 2022a], subprime financial and economic crisis or in the case of natural disasters [Gargani, 2022a and b; Akter, 2023; Van der Borght and Pallares-Barbera, 2024].

The concentration of CO2 in the atmosphere is increasing along with energy production (Figure 1) [Huang and Lixin Tian, 2021], particularly with regard to hydrocarbon use [IPCC, 2013. Energy is produced using a variety of resources (oil, gas, nuclear, coal, water storage for hydroelectricity, wind, solar energy, geothermal energy, tidal power plant, etc.) [Radanne and Puiseux, 1989]. The production of energy is criticized for several reasons including: (i) pollution and greenhouse gas emissions, (ii) nuclear risks, (iii) geopolitical dependence, (iv) resources scarcity, (v) resources costs, (vi) the production of socio-economic pathologies [Latouche S., 2004; Bonneuil and Fressoz, 2013; Sovacool et al., 2022; Blondeel et al., 2024].

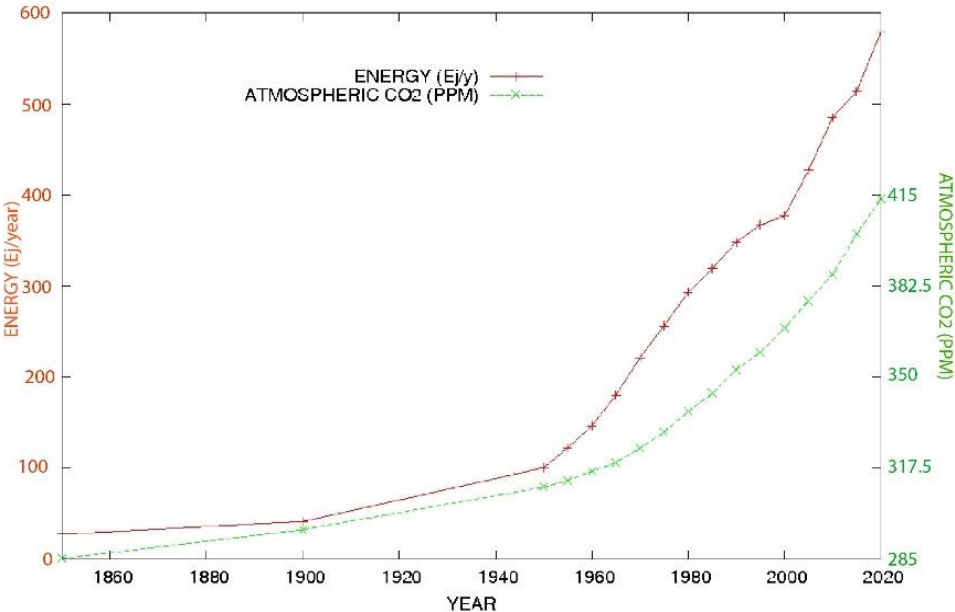

**FIGURE 1**: *Energy production and atmospheric CO₂ evolution (ppm) with time. Data source: https://static-content.springer.com/esm/art%3A10.1038%2Fs43247-020-00029-y/MediaObjects/43247_2020_29_MOESM1_ESM.pdf [Bolt et al., 2018;]*

The various energy sources have different effects on atmospheric greenhouse gases and environment. Renewable energy has a lower impact because little or no carbon dioxide emissions are directly produced, but is also criticized because it does not substitute for other energy sources [Fressoz, 2024]: According to Fressoz (2024) new energies increase energy production and resource consumption, while not reducing it. The type of energy used is frequently a political choice: (i) depending on geopolitical strategies, (ii) depending on health and environmental concerns, (iii) depending on economic and social strategies [Radanne and Puiseux, 1989].

The Earth's climate changes and has consequences on both a global and local scale. This study will investigate the local effetcs of climate warming will be investigated. First, extreme events occur on a local or regional scale. Second, because the monitoring on local or regional level is more accurate and can be used as a survey. Third, because investigating multiple territories permit to test a method. Four, local initiatives can contributes to global impact [Petrovics et al., 2024]. To characterize the effect of climate impacts (precipitation, temperature, extreme events) on energy consumption, this study focuses on areas where extreme events and relatively high temperatures have been observed. Small tropical islands are particularly sensitive and vulnerable [Prina et al., 2021]. They are hyper-reactive laboratories for global warming. Studying these territories provides an opportunity of carrying out a detailed analysis of the interactions between a territory and its environment. More specifically, the small tropical islands of La Réunion, Guadeloupe, Martinique, Saint-Martin, Saint-Barthélemy, Mayotte, French Polynesia and Nouvelle-Calédonie will be studied using their energy consumption to assess their vulnerability to climate change.

Energy has been shown to influence climate, but climate can also influence energy production and consumption [Roman et al., 2019; Gargani, 2022b]. A better understanding of how climate affects energy consumption on these islands is used to assess their vulnerability to climate change. The interactions between society and the environment will be investigated by monitoring energy production over time. More specifically, how extreme climatic events and temperature increases affect consumption will be investigated. Is there a temperature variation threshold above which there is an impact on electrical energy consumption? At what threshold (precipitation/wind) do cyclones leave significant traces in energy production? Is energy consumption a useful indicator for determining the amplitude of an event (anthropic or climatic)? Can the effects of global warming on energy consumption be anticipated?

## 2. Local context and method

### 2.1. Method

Isolated environments disconnected from the rest of the energy network were chosen in order to obtain the most comprehensive energy assessment possible. Characterizing variations in energy consumption and their correlation with parameters such as demographic growth, wealth variation and climatic changes, all other things being equal, is more accurate in small isolated territories than on a global scale.

Primary energy production is indicative of numerous activities. Primary energy production provides information on transport, industrial production and social activities. Individual and collective practices such as the use of air conditioning, new professional and domestic appliances, or an increase in the number of electric cars, can influence the evolution of electricity production. The variation of electric energy consumption permits to characterize variations in activities [Hall and Klitgaard, 2011; Li et al., 2018; Jones et al., 2015].

Studying small tropical islands has some benefits. There are few exports due to low internal production. However, there are number of limitations. Imports of primary energy are only counted for oil or gas, but not for the energy required to produce imported products, resulting in an understimates of the total energy consumption required to maintain the way of life on these islands is underestimated. This is because the total energy consumed in these territories as a result of imported production (food, household appliances, automobiles, etc.) is not calculated using primary energy consumption.

Furthermore, it is difficult to obtain accurate time steps for primary energy production. However, many activities require electricity. This study will analyze electricity consumption and not just primary energy production. In terms of energy, the data used in this study are mainly obtained from EDF SEI [EDF, 2015, 2018a, 2018b, 2023a, 2023b, 2023c, 2023d; Ecoconcept Caraibes, 2018], the La Réunion Energy Observatory (Observatoire de l'Energie de La Réunion)[Observatoire de l'énergie de la Réunion, 2011, 2013a, 2013b; observatoire Polynésien de l'énergie, 2017; Observatoire de la transition écologique et énergétique, 2022] and data collected by IEDOM or IEOM (French institutes specialized in the synthesis of the French overseas territories' economy).

In order to study the influence of the different factors (*i.e.* number of inhabitants, GDP/inhabitants, temperature variation, cyclone occurrence) on energy production and consumption, graphical trends will be presented (section 3). The relationships between parameters were described using trends rather than statistical correlation. A qualitative analysis is performed and, then, the interpretation are discussed. Finally, the identification of causal relationships between these parameters is suggested.

A comparison between the small tropical islands is performed to decipher the influence of the number of inhabitants in sections 3.1 and 3.2. An estimation of the mean value for the period 2010-2020 of: (i) the primary energy consumption, (ii) the electricity production.

Comparing changes in electricity production is a useful method to assess the impact of a specific events (climatic, cultural, and political) on the socioeconomic world. To compare the electricity production on the islands of La Réunion, Saint-Martin and Mayotte, the energy production was normalized by the number of inhabitants for each islands, as well as the amount of electricity produced in 2010 (section 3.3). This second normalization enables a more accurate graphical representation and an easier comparison.

The impact of rising temperatures on electricity production was estimated by comparing temperature variations to electricity production (section 3.4). The data were obtained from Météo France (*https://meteofrance.re/fr*) for each month over a 6-year period. The mean monthly temperature in La Réunion from 2017 to 2022 is calculated. The temperature excess above 26°C (the coldest monthly temperature) in La Réunin Island was calculated for each month. The coldest month on La Réunion is June.

Furthermore, using data from the La Réunion Energy Observatory [Observatoire de l'énergie de La Réunion; *https://oer.spl-horizonreunion.com/*], the average amount of additional electrical energy for each month from 2017 to 2022 was estimated in comparison to the one in June. Electricity production is used instead of primary energy production because it is more sensitive to temperature fluctuations. For each month, the temperature excess is correlated with the excess of electricity production. The error represents the standard deviation of temperature and electricity production.

Deviations from the average yearly temperature are calculated by estimating the average temperature value from 1991 to 2020 (*i.e.* 30 years). The average decadal (30-year) value is then subtracted from the mean yearly temperature. The annual deviation from the average electricity production is calculated by estimating the mean decadal electricity production from 1991 to 2020. The difference between annual electricity production and the average decadal (30-year) value is calculated. Finally, these two deviations from the mean for temperature and electricity production were plotted on a graph spanning the years 2000 to 2022.

### *2.2.Small tropical island context*

This study focuses on: (i) islands with independent electricity production that are disconnected from other territories, (ii) tropical environments with relatively high mean temperatures (>20°C) and the possibility of extreme events such as cyclones, and (iii) territories administered by France, because they have a relatively similar data collection process.

More precisely, the territories studied are Saint-Martin, Saint-Barthélemy, Guadeloupe, Martinique, La Réunion, Mayotte, French Polynesia and Nouvelle-Calédonie. While Saint-Martin [Pasquon et al., 2019, 2022a; IEDOM, 2020b, 2023b], Saint-Barthélemy [Chardon and Hartog, 1995; Pasquon et al., 2022b; IEDOM 2012, 2020a, 2021a, 2023e], Guadeloupe [Artelia, 2020; IEDOM, 2023a] and Martinique [IEDOM, 2023b; Observatoire territorial de la transition écologique et énergétique, 2022] are located in the Caribbean, La Réunion [ARER, 2010; IEDOM, 2014; IEOM, 2023a; Préfet de La Réunion, 2019] and Mayotte [Hachimi Alaoui et al., 2013; Deves et al., 2022; IEDOM, 2015, 2023c; Préfet de Mayotte, 2016; Tsimonda, 2023] are located in the Indian Ocean and French Polynesia [IEDOM, 2010; IEOM, 2023d; Meyer T., 2021; Observatoire Polynésien de l'énergie, 2017] as well as the Nouvelle-Calédonie Island [IEOM, 2013, 2016, 2019, 2020, 2023b] are located in the Pacific Ocean.

Except for the Nouvelle-Calédonie Island, all of these islands are volcanic in origin [Gargani, 2020; Gargani 2022c; Gargani, 2023; Gargani 2024]. Volcanic activity can be very old (>10 Ma for Saint-Martin and Saint-Barthélemy), relatively young (<1 Ma, French Polynesia), or consistently active in the last centuries (Guadeloupe, Martinique, Mayotte, La Réunion). Present or recent volcanic activity could be favorable to produce geothermal energy.

These islands: (i) are tropical, (ii) have an economy based on tourism since the 1980s, (iii) have experienced a significant increase in their GDP –Gross Domestic Product– in recent decades, (iv) have experienced significant demographic growth over the last decades, but (v) have very different GDP/inhabitants, (vi) have not the same number of inhabitants (Table 1), and (vii) have no industrial production activity, except for the Nouvelle-Calédonie Island [ARER, 2010; IEDOM, 2010, 2012, 2014, 2015, 2020a, 2020b, 2021a, 2021b, 2023a, 2023b, 2023c, 2023d, 2023e; IEOM, 2016, 2019, 2020, 2022, 2023a, 2023b, 2023c; Préfet de La Réunion, 2019; Préfet de Mayotte, 2016]. Although there are administered by France and were once part of the French colonial empire, they do not currently have the same level of autonomy, at the present time. Political debates over increased autonomy or independence may be conflictual in these islands.

*Table 1*: *Socio-economic features of the studied islands. NC= Nouvelle-Calédonie. Data on population and GDP come from INSEE (https://www.insee.fr/), but also from IEDOM [IEDOM, 2010, 2012, 2014, 2015, 2020a, 2020b, 2021a, 2021b, 2023a, 2023b, 2023c, 2023d, 2023e] and IEOM [IEOM, 2016, 2019, 2020, 2022, 2023a, 2023b, 2023c].*

| Island | GDP/inhabitant (euros) | Mean population (2010-2020) | Electricity/yr (mean 2010-2020, GWh) | Electricity/yr/inhabitant |
|---|---|---|---|---|
| La Réunion | 22359 (in 2018) | 846062 | 2876.7 | 0.00340 |
| Mayotte | 10600 (in 2021) | 238409 | 303.2 | 0.00127 |
| Saint-Martin | 16572 (in 2014) | 34965 | 187.3 | 0.00535 |
| Saint-Barthelemy | 38994 (in 2014) | 9628 | 124.4 | 0.0129 |
| French Polynesia | 18572 | 279679 | 685.1 | 0.00245 |
| Guadeloupe | 23449 | 396286 | 1730.6 | 0.00436 |
| NC (Metallurgy) | 31584 | 2699184 | 3062.9 | 0.01138 |
| NC (No Metal.) | - | 2699184 | 799 | 0.00297 |
| Martinique | 25604 | 376505 | 1562.8 | 0.00415 |

The climate of the tropical islands investigated in this study is divided into two seasons: dry and wet. During the wet season, cyclones and storms are expected to occur. The mean temperatures in these islands ranges from 22°C to 35°C. Weather data is provided by Météo France *(https://meteofrance.re/fr)*. Climatic warming trends and rising sea levels indicate that these islands are becoming more vulnerable to extreme hydro-climatic events. The question of their adaptation and its modalities is controversial.

When volcanism, relief, and rainfall are favorable, some of these islands have developed renewable energy production, which can account for up to 30% of total electricity production, such as in La Réunion Island with hydroelectric energy and in Guadeloupe with geothermal energy. However, the energy dependence of these islands is very significant (>80%) in relation to the primary energy consumed, based on gas and oil consumption [Syndicat des énergies renouvelables, 2018; INSEE, 2024] (INSEE, https://www.insee.fr/).

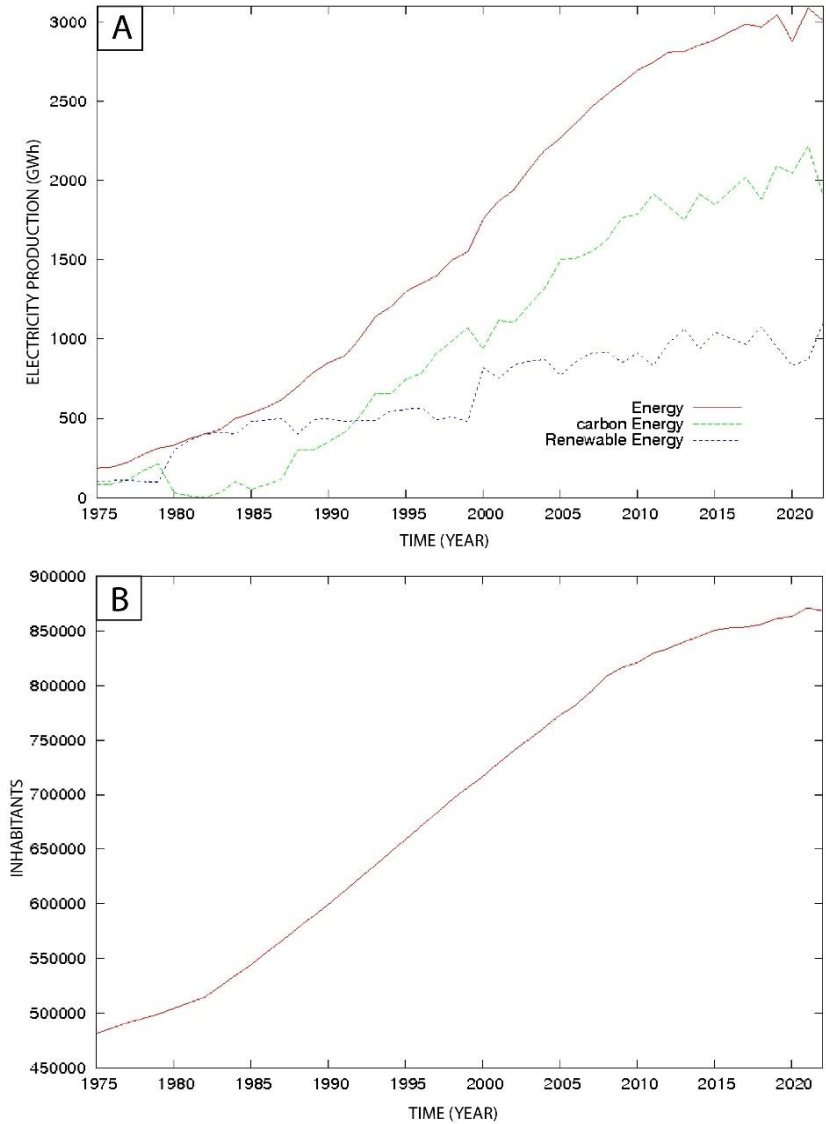

***FIGURE 2**: (A) Annual electricity production in La Réunion Island between 1975 and 2022. Renewable energy in blue (mainly hydroelectricity) and carbon electricity in green (oil and gas), (B) Evolution of the number of inhabitant with time. Data source : the Observatoire de l'énergie de La Réunion (https://oer.spl-horizonreunion.com/) and INSEE (https://www.insee.fr/).*

Between 1975 and 2020, electricity production on La Réunion Island increased steadily (Figure 2A). Electricity production accounts for 15–18% of total energy consumption on the island of La Réunion. Electricity production growth accelerates from 1975 to 2000, but then slows between 2000 and 2020 (Figure 2A). Electricity consumption has increased slightly in comparison to total energy consumption over the last few decades.

Contemporaneously, there is an increase in the population, which appears to follow the same pattern (Figure 2B). From 1980 to 2000, the population growth rate has accelerated. Following 2000, the variation in the number of inhabitants slows down.

Renewable energies increased gradually from 1975 to 2020. More specifically, several growth pulses in renewable energy production can be observed in 1975, 2000, and 2011. However, the amount of electricity generated from carbon-based energy is also increasing. Increasing consumption of carbon-based energy is observed despite the negative impact on the environment [IPCC, 2013]. Some annual variations in carbon-based energy have been decided to compensate for declining renewable energy productions beginning in 2021.

***Table 2*** *: La Réunion Island energy consumption and production. Data source : Observatoire de l'énergie de La Réunion (https://oer.spl-horizonreunion.com/).*

| Year | Fossil Energy (ktep) | Renewable Energy (ktep) | Total Primary Energy (GWh) | Electricity Production (GWh) |
|---|---|---|---|---|
| 2000 | 816.1 | 156.7 | 11314 | 1758.1 |
| 2005 | 1043.8 | 149 | 13872 | 2270.6 |
| 2010 | 1217 | 174.2 | 16180 | 2698.8 |
| 2015 | 1214.8 | 196.5 | 16415 | 2890.5 |
| 2017 | 1277.5 | 189.5 | 17061 | 2986.0 |
| 2018 | 1256 | 186.7 | 16779 | 2969.7 |
| 2020 | 1191 | 177.7 | 15918 | 2878.5 |
| 2022 | 1227.4 | 203.2 | 16638 | 3012.0 |

### 3.1.2 General case

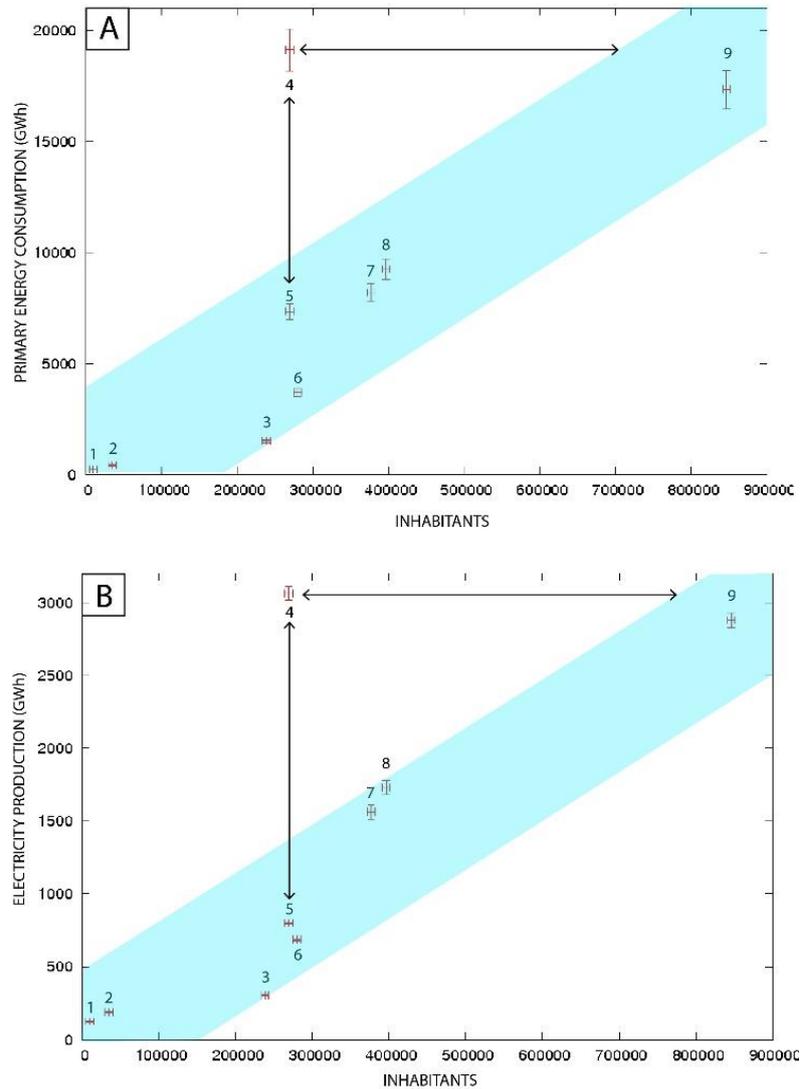

**FIGURE 3**: *(A) Influence of the number of inhabitants on annual primary energy consumption in Islands, (B) Influence of the number of inhabitants on annual electricity production. From the smaller inhabitants number to the more: 1-Saint-Barthelemy, 2-Saint-Martin, 3-Mayotte, 4-Nouvelle-Calédonie with metallurgy, 5-Nouvelle-Calédonie without metallurgy, 6-French Polynesia, 7-Martinique, 8-Guadeloupe, 9-La Réunion).*

On small tropical islands, the larger the population, all other factors being equal (*i.e. ceteris paribus*), the higher the energy consumption (Figure 3). The energy production capacity is proportionate to the number of inhabitants. The case of the Nouvelle-Calédonie Island shows that local ore extraction and production (extraction of nickel and metallurgical production) require a significant amount of energy. The inhabitants of Nouvelle-Calédonie Island consume as much energy as an "equivalent" island with over 800,000 inhabitants, nearly three times the actual population. The energy used for metallurgy purpose is included in the Nouvelle-Calédonie Island energy balance, but not in the energy balances of the territories where metallurgic product will be used. Including energy consumption associated with the production of imported products would significantly increase energy consumption on the small tropical islands. Nonetheless, this bias is systematic, as no production occurs on the other small tropical islands studied here.

Primary energy consumption and electricity production in La Réunion Island are consistent with energy and electricity production and consumption on other small tropical islands. The influence of the number of inhabitants and the increase in the number of inhabitants must be considered when interpreting variations in electricity consumption. To avoid misinterpretation of the variation in electricity consumption, the indicator of electricity consumption per capita will be used in the following sections.

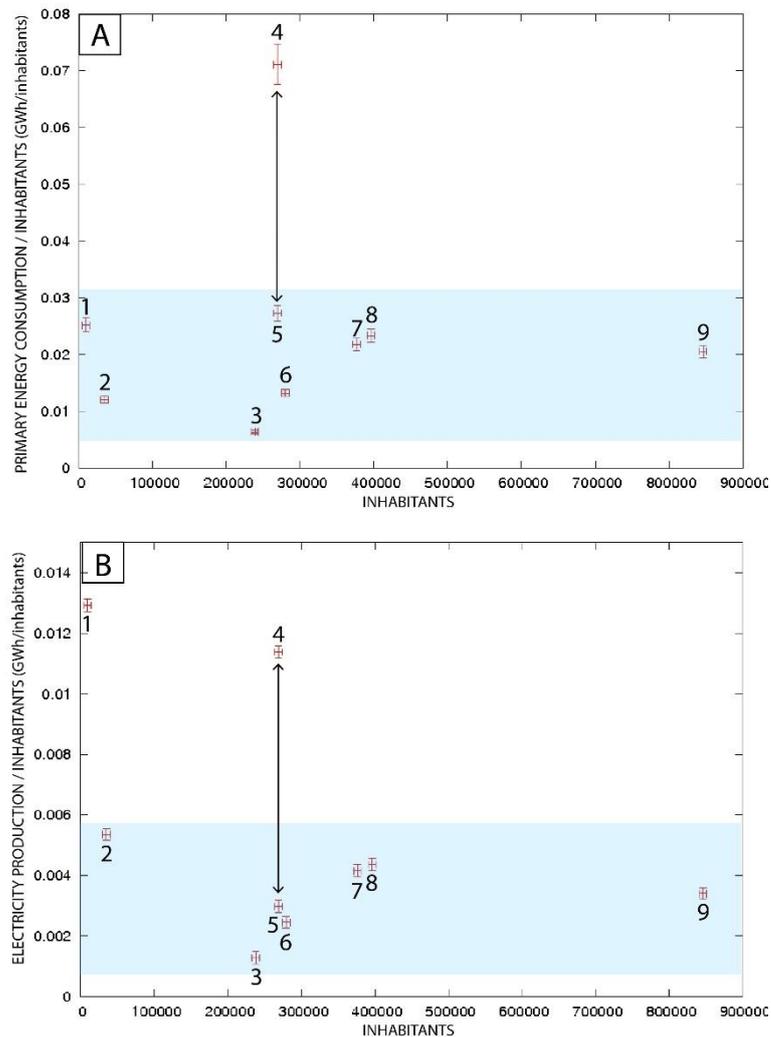

***Figure 4****: Influence of inhabitant's number on primary energy consumption per inhabitants. 1-Saint-Barthelemy, 2-Saint-Martin, 3-Mayotte, 4-Nouvelle-Calédonie with metallurgy, 5-New Caledonia without metallurgy, 6-Polynesia, 7-Matinique, 8-Guadeloupe, 9-La Réunion. Number of inhabitant between 2010 and 2020 (*https://www.insee.fr/*). Mean electricity production between 2010 and 2020 from EDF, Observatoire de l'énergie de La Réunion, IEDOM and IEOM.*

The population increase has no significant effect on primary energy consumption per capita in these islands (<1 million of inhabitants). No trend is observed (Figure 4). There is no apparent increase in economic efficiency when the population of the small tropical islands grows.

It can be observed again that the Nouvelle-Calédonie Island has higher per capita primary energy consumption and electricity production than expected. Overproduction and consumption are caused by nickel extraction and transformation (*i.e.* metallurgy). In Saint-Barthelemy the electricity production per capita is higher than in other small tropical islands. At the opposite, electricity production in Mayotte is relatively low. Section 3.3.3 will look into the explanations for these features.

Normalizing energy consumption and electricity production by the number of inhabitants appears to be able to mitigate a potential bias when comparing small tropical islands. Population growth may have an impact on economic growth [Heady and Hodge, 2009], but it appears to be negligible in small tropical islands based on electricity production (Figure 4). In contrast, this indicator (electricity production/inhabitants) shows no negative effect from population growth.

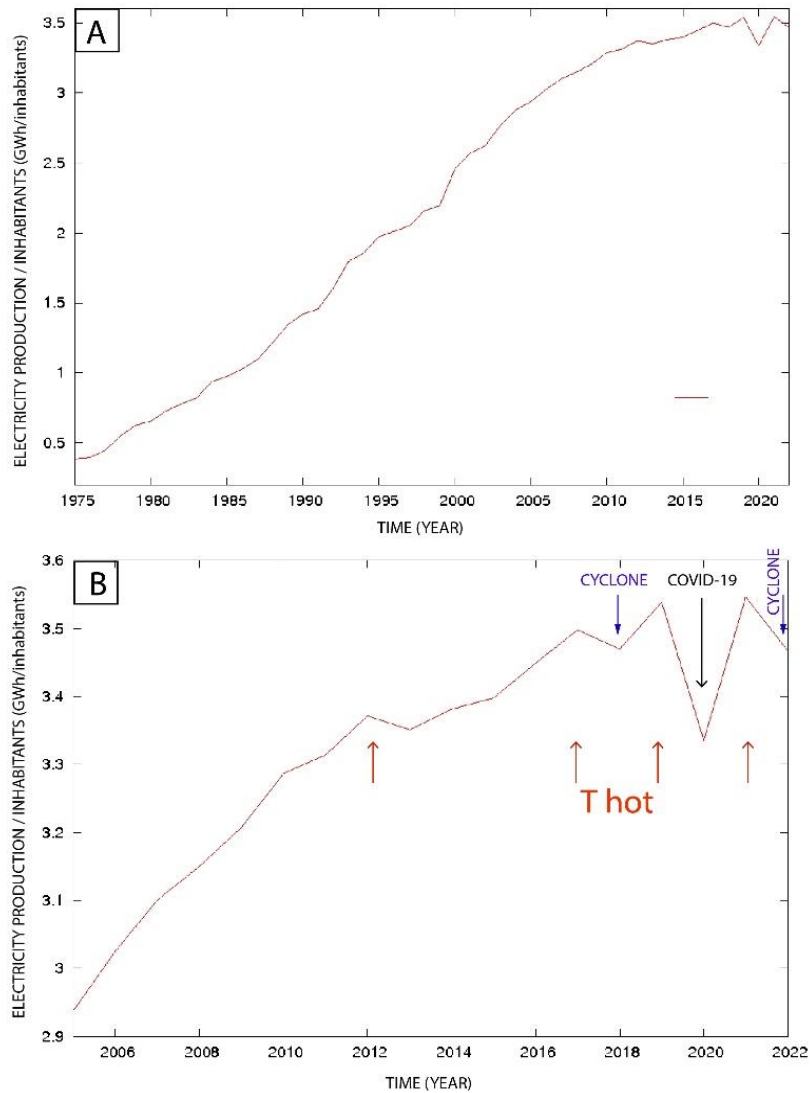

**FIGURE 5**: *(A) Annual electricity production in La Réunion Island between 1975 and 2022 normalized by the number of inhabitants. (B) Annual electricity production in La Réunion Island between 2007 and 2022 normalized by the number of inhabitants Data sources: INSEE (*https://www.insee.fr/*) for the number of inhabitants and EDF for the electricity production [Observatoire de l'énergie de la Réunion, 2011, 2013a, 2013b; Préfet de La Réunion, 2019].*

When electricity production is normalized for the number of inhabitants, energy production per capita increases from 1975 to 2022 (Figure 5). The increase in energy consumption (Figure 2A) is not only due to the population growth, but is also caused by other factors. Let us now investigate the potential causes of energy consumption increases.

New lifestyles and businesses may influence activities and consumption. Tourism has grown in many countries worldwide since the 1980s, including on tropical islands such as Saint-Barthélemy and Saint-Martin [Chardon and Hartog, 1975; Pasquon et al., 2022a]. The development of this economic feature modified the activities on La Réunion Island, favoring unemployment reduction. During the last decades, the GDP increased significantly in relation with economic growth in La Réunion as well as in others small tropical islands.

The general growth in electricity production is of around 10% during the last period (2010-2022), which corresponds to a growth of around 1% per year. Nevertheless, this growth is smaller than previously. Let us investigate if the slowdown in energy consumption (even if the growth rate remains > 0) since 2010 is due to: (i) an economic slowdown, (ii) a climatic effect, or (iii) an increased efficiency in the use of energy.

The role of wealth and socio-economic crisis on long-term evolution of electricity consumption will be studied in section 3.3.2 and 3.3.3, respectively. The impact of climate on electricity consumption will be analyzed in section 3.3.4. The efficiency in the use of energy will be discussed in the discussion section (section 4.).

### 3.3.2. Influence of wealth

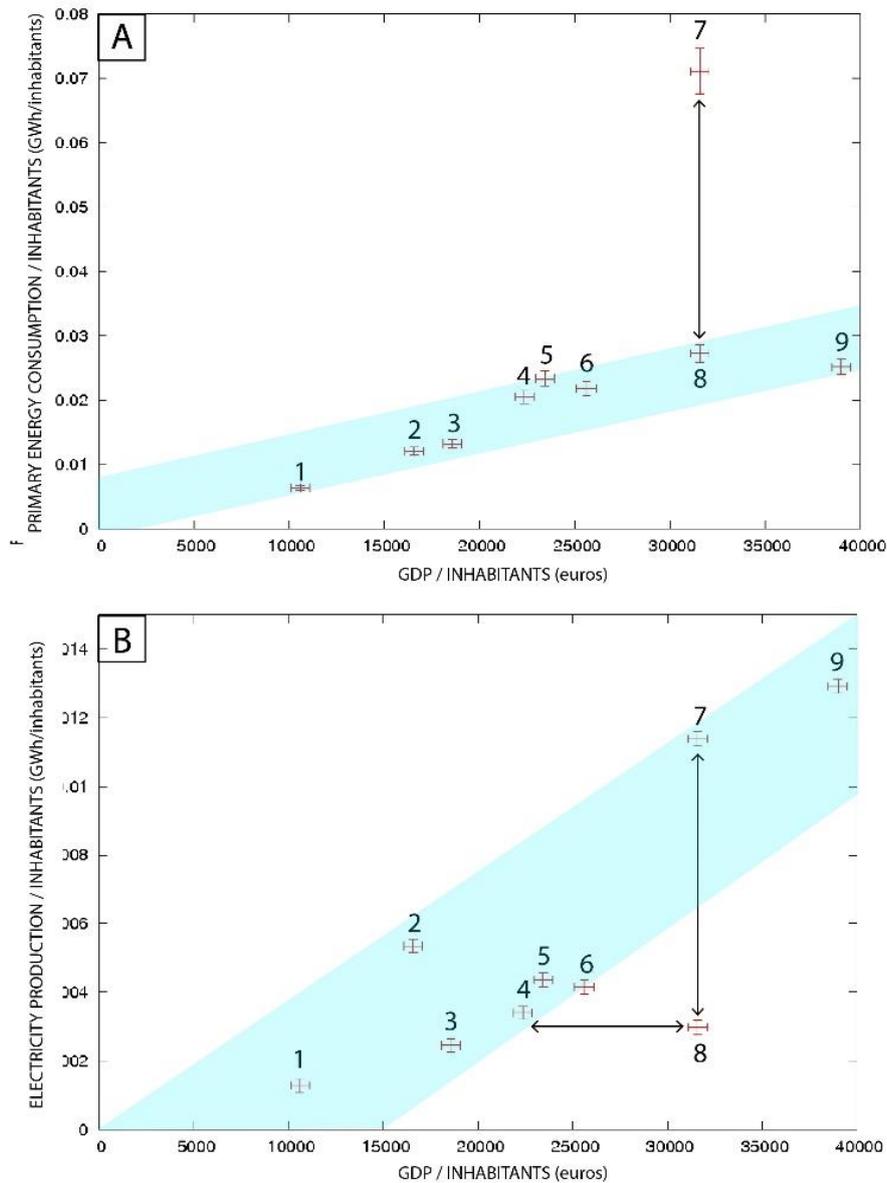

***FIGURE 6***: *Influence of wealth on energy consumption. (A) Influence of the GDP/inhabitant on primary energy consumption per inhabitants in small tropical islands, (B) Influence of the GDP/inhabitants on electricity production per inhabitants. From the smaller GDP/inhabitant to the higher:1-Mayotte, 2-Saint-Martin, 3-French Polynesia, 4-La Réunion, 5-Guadeloupe, 6-Martinique, 7-Nouvelle-Calédonie with metallurgy, 8- Nouvelle-Calédonie without metallurgy, 9-Saint-Barthelemy).*

To better understand the evolution of electricity production per capita, it is necessary to consider the inhabitants' wealth and evolution. The greater the GDP per capita, the higher the annual energy consumption per capita (Figure 6A). The causes of the wealth increase are beyond the aim of this study. In this study, the wealth is described using the GDP per capita. This indicator has been criticized because it fails to characterize accurately the inequality within a territory (health, education, etc.) [Piketty, 2013; Stiglitz et al., 2018].

The greater the GDP per capita (i.e.), the more the people have the possibility to have expensive social and cultural activities that increase electricity consumption. For example, the capacity to have a boat depends on the incomes. Furthermore, the higher the GDP/inhabitant, the larger the houses and the more they have a swimming pool [Pasquon et al., 2022a]. In this case, energy consumption could increase when people heat their swimming pool or use air conditioner in larger houses. The increase of the GDP per capita could be associated with various activities and lifestyles. Use of new connected technologies (electric cars, computers, smartphones, air conditioners, etc.) increase the electricity consumption. However, f the new inhabitants are very poor, the mean GDP/inhabitant decreases, but the electricity consumption per capita could be stable.

Electricity production is not always consumed directly by residents, but could be consumed by other activities such as metallurgy in Nouvelle-Calédonie or tourism (in Saint-Martin). Significant amount of energy in are produced on Nouvelle-Calédonie Island for nickel extraction and metallurgical production for export, rather than for local use. Energy consumption of the mineral mining industry is significant [Aramendia et al., 2023]. The extraction of nickel and the transformation of the ore by the metallurgical industry enriched the population in a very inhomogeneous way. In Nouvelle-Calédonie, the 10% with higher incomes are 7.1 times richer than the 10% with lower incomes, compared to 3.5 times in France [Salaün and Trépied, 2024]. When inequality is very high, the mean wealth obtained using the GDP per inhabitant, is insufficient to characterize accurately a territory.

The electricity consumption excluding metallurgy in New Caledonia is relatively low compared to the GDP per capita. The GDP per capita without metallurgy is less than 25 keuros per inhabitants, according to the graph (Figure 6B). The extraction of raw materials (nickel) and production for export increases the GDP/inhabitants, it does not seem to generate lifestyles comparable to those found on equivalent tropical islands in New Caledonia. Saint-Martin's energy production is also slightly higher than expected in terms of GDP per capita.

On the one hand, wealth decline could be attributed to: (i) socioeconomic crisis, (ii) epidemic crisis, and (iii) climatic crisis. On the other hand, wealth growth in these islands could be attributed to: (i) new public investment, (ii) new economic activity development, such as tourism, (iii) new technological development, or (iv) new collective organization.

### 3.3.3. Influence of crises on electricity production

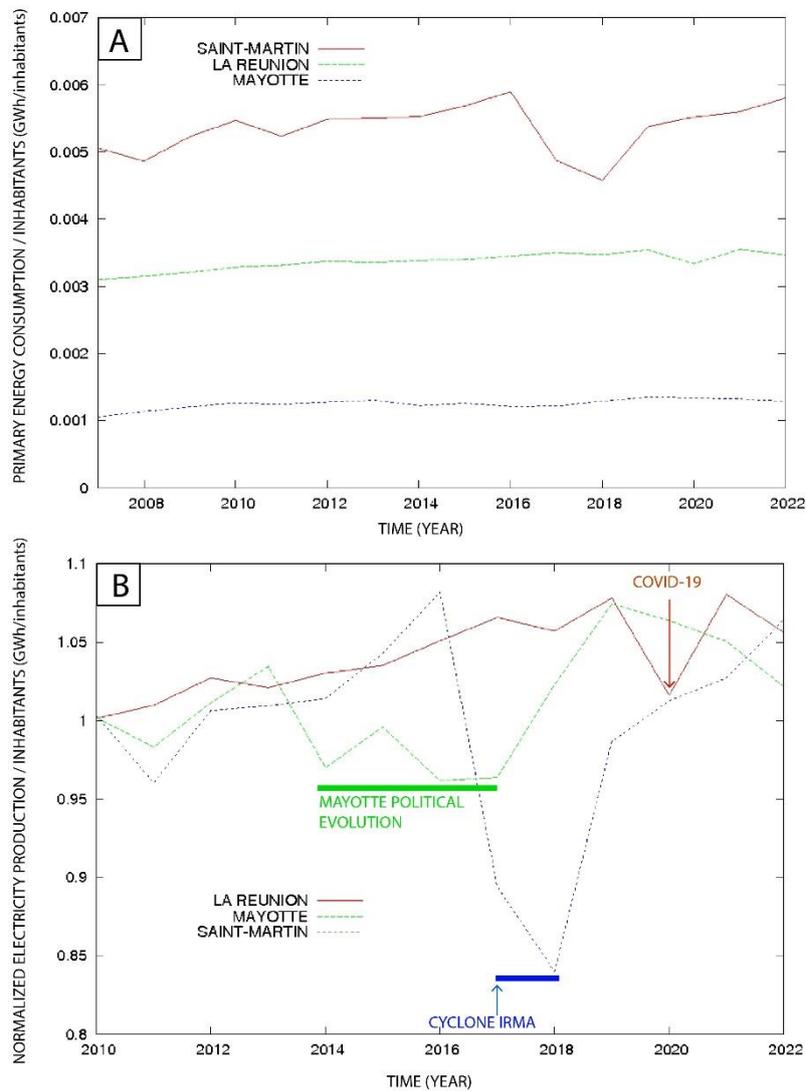

***FIGURE 7****: (A) Annual electricity production in Mayotte, Saint-Martin and La Réunion islands normalized by the number of inhabitants, (B) Annual electricity production in Mayotte, Saint-Martin and La Réunion islands between 2010 and 2022 normalized by the number of inhabitants and the electricity production in 2010. Data sources: INSEE (*https://www.insee.fr/*) [INSEE, 2024], IEDOM [IEDOM, 2014, 2015, 2019, 2020, 2023c, 2013e] and IEOM [IEOM, 2023b,] and Observatoire de l'Energie de La Réunion (https://oer.spl-horizonreunion.com/) [Observatoire de l'énergie de La Réunion, 2011, 2013, 2013].*

Small tropical islands may be vulnerable to natural hazards (cyclones, heavy precipitation, drought, marine submersion, landslides, erosion, and earthquakes), but they are also impacted by anthropogenic events (economic development, technological development, crisis economic, demographic development, migration, health crisis).

Each island is impacted by various events that are not necessary similar. To compare electricity production between different islands, such as La Réunion, Mayotte, and Saint-Martin, it is necessary to normalize the indicator by the number of inhabitants. Indeed, as previously demonstrated, the number of inhabitants has a significant influence on electricity and energy consumption. First, the number of inhabitants is used to normalize electricity production, as La Réunion has more people than Mayotte and Saint-Martin. The normalized electricity production per capita in Saint-Martin is higher than in La Réunion and Mayotte (Figure 7A), making comparison difficult.

To make a more detailed comparison of the evolution of electricity consumption on these islands, it may be necessary to normalize the indicator by the electricity production of a reference year. Electricity production was normalized relative to consumption in 2010. The economic growth in small tropical islands has been impacted in 2008-2009-2010 by the consequences of the subprime crisis that began in the United States. This impact can be observed in Saint-Martin and Saint-Barthélemy [Gargani, 2022a]. Tourist arrivals, unemployment and more generally economic growth have all been impacted [INSEE, 2014; https://www.insee.fr/fr/statistiques/128527 8]. When electricity production increases is very high, the record of crisis cannot always be observed. The Chikungunya epidemic on La Réunion Island in 2006 had a significant economic impact, but had no significant impact on electricity production because during the epidemic crisis, the residents stayed at home to avoid mosquito risk and consumed electricity as they did during all the others wet seasons.

The effect of Covid-19 can also be observed on annual electricity production of La Réunion Island. In Mayotte, the annual electricity production decreased in 2020 and after. This observation is in agreement with observations in Europe [Halbrügge et al., 2021] and elsewhere [Jiand et al., 2021; Navon et al., 2021].

Even if it is not clearly observed in the annual electricity production of Saint-Martin, Covid-19 has significant impact on the island's economic and social activities in Saint-Martin in 2020 [Gargani, 2022a; IEDOM, 2023b]. Hurricane Irma (2017) [Cangialosi et al., 2018; Jouannic et al., 2020] had a significant impact on annual electricity production in Saint-Martin, not only in 2017 but also subsequently. The impacts of this cyclone on the economy of the island were still observable in 2020 when the pandemic crisis occurred. Consequently, in 2020, the ongoing recovery of the Saint-Martin Island following Irma's destruction superimposed the impact of the Covid-19 on social and economic activity. Covid-19 had a significant negative impact on tourism, the main economic resource of the Saint-Martin Island [IEDOM, 2023b], but its impact was more difficult to distinguish from that of Hurricane Irma on electricity consumption. Tourist arrivals at the airports and harbor were significantly reduced. The impact of Covid-19 on electricity production in Saint-Martin was partially hidden by the contemporaneous recovery of the economic activity following Irma.

Hurricane Irma caused a population departure in Saint-Martin (7000-8000 people left), and the island's population is still approximately 3000-4000 lower than it was prior to the hurricane. The population decrease may explain part of the electricity production decrease in Saint-Martin after 2017 [Gargani, 2022a].

Social and economic activities evolutions could also explain the decrease in electricity production observed in Mayotte between 2014 and 2018 (Figure 7B). In Mayotte, there was a change in 2014 regarding: (i) tax laws, (ii) the code of entry and stay for foreigners and the right to asylum. Concerning the first point, Mayotte was designated as an "outermost region" of the European Union in 2014. As a result, new rules were applied. French laws, including new financial taxation, were applied in ways that had not previously been done. Concerning the second point, it was suggested that migrants arrived in Mayotte and that the number of inhabitants increased until 400000 (French Home Minister; gendarmerie.interieur.gouv.fr, 2021). Nevertheless, this was considered experts from the French statistic institute considered this to be an overestimation (*blog.insee.fr/mayotte-census-adapted-to-non-standard-population*)

### 3.4 Climatic impact

### 3.4.1. Seasonal influence and extreme hydro-meteorological events

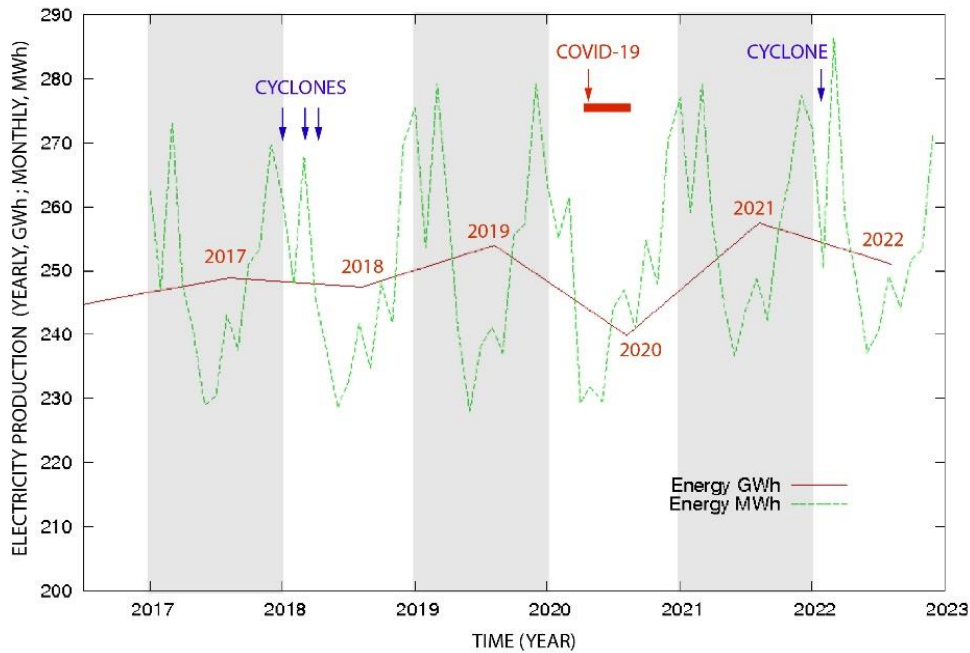

***FIGURE 8***: *(A) Monthly Electricity production in La Réunion Island between 2017 and 2022 (in green) and annual electricity production (in red). Data source: Observatoire de l'Energie de La Réunion (https://oer.spl-horizonreunion.com/).*

Monthly data show that electricity production varies seasonally (Figure 8. It also varies throughout the day, but that is not the focus of this study. The variations are caused by cyclical activities, specifically socioeconomic activities. In particular, electricity production is greater during the Southern Hemisphere summer (November-December-January-February-March) than during the southern winter (May-June-July-August-September). Every year, there are two peaks, centered in January and March.

As previously mentioned, extreme events can cause significant damages [Kishore et al., 2018; Roman et al., 2019; Howell and Elliott, 2019; Blaikie et al., 1994; Cutter, 1996; Cutter et al., 2003; Pichler and Striessnig, 2013; Rubin and Rossing, 2012] and significant changes in energy consumption as seen in the case of Hurricane Irma in Saint-Martin and in Saint-Barthélemy. The primary causes of energy consumption decrease during and after extreme events are: (i) destruction or damage to power plants, (ii) destruction or damage to electricity networks, (iii) destruction or damage to infrastructures, (iv) migration and population decrease, and (v) economic and social activity slowdown [Roman et al., 2019; Gargani, 2022a and b; Der Sarkissian et al., 2021 and 2022].

Initially, the destruction of electricity production plants, as well as energy distribution when there is complete or partial destruction of the electricity network [Roman et al., 2019; Der Sarkissian et al., 2021 and 2022]. Finally, through the destruction of other infrastructure, such buildings and transportation vehicles, has a significant economic impact. A blackout can have severe health consequences [Roman et al., 2019]. Territorial recovery can take years to return to a situation similar to the original when especially one, when the damages are severe [Gargani, 2022b].

In 2018, La Réunion island was impacted by three cyclones during 11 days (Figure 8).

In 2022, La Réunion Island was impacted by cyclone Batsirai during 4 days (Table 3). Heavy rains fell during these cyclones were recorded (Table 3, Météo France data source; *https://meteofrance.re/fr*). Nevertheless, during the same events, winds did not significantly damage infrastructures or buildings. Consequently, electricity consumption decreased, but not as significantly as it had during Hurricane Irma. One of the climates related effect is a decrease in electricity production.

*Table 3*: *Main climatic events during 2017-2022. Data source: Météo France (https://meteofrance.re/fr).*

| Year | Extreme events | Precipitation maximum (mm) | Mean annual precipitation variation (normalized by the mean value obtained between 1981-2010) | Mean annual temperature variation (normalized by the mean value obtained between 1981-2010) |
|------|----------------|----------------------------|-----------------------------------------------------------------------------------------------|---------------------------------------------------------------------------------------------|
| 2022 | Cyclone Batsirai | 584 mm in 4 days, February | + 5% | + 0.1°C |
| 2021 | | | - 5% | + 0.65°C |
| 2020 | | | -25% | + 0.2°C |
| 2019 | | | -21% | + 1.2°C |
| 2018 | Cyclone Ava Cyclone Dumazile | 558 mm in 6 days, January 541 mm in 4 days, Mars 176 mm in 1 hour, April | + 40% | + 0.65°C |
| 2017 | | | -8% | + 0.9°C |

### 3.4.2 Influence of temperature

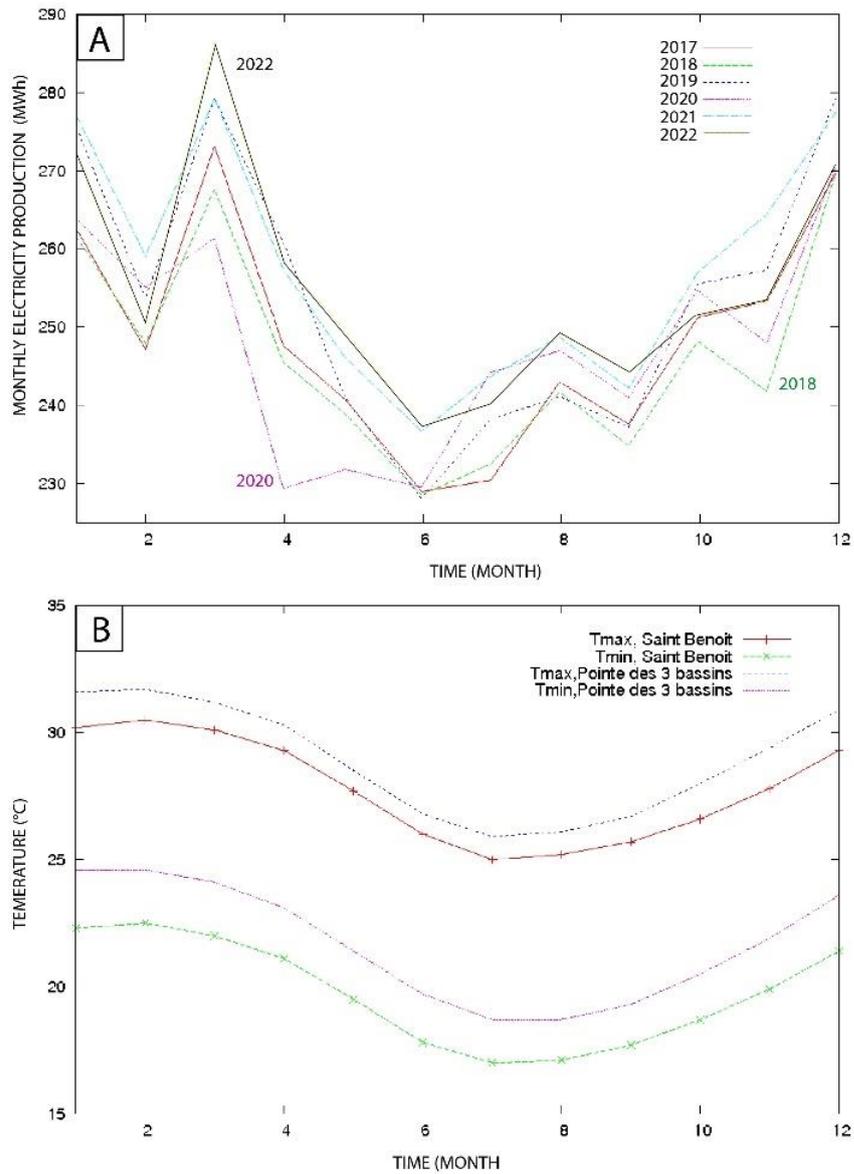

**FIGURE 9**: *(A) Monthly Electricity production in La Réunion Island between 2017 and 2022, (B) Mean monthly precipitation, (C) Mean monthly temperature for maximum and minimum temperature in Saint Benoit (north-east of La Réunion island) and Pointe des 3 bassins (west of La Réunion island). Data source from Météo France (https://meteofrance.re/fr) for precipitation and temperature. Data are collected by IEOM and DIMENC for electricity production.*

Seasonal increase of electricity consumption occurred during southern summer in La Réunion Island, when the precipitations and the temperatures are higher (Figure 9). In 2018, 2020 and 2022, a small decrease of the electricity production can be observed in comparison to 2017, 2019 and 2021 (Figure 8). As previously described, COVID-19 had a significant impact on energy consumption in 2020. The occurrence of cyclones in 2018 and 2022 decreased energy production slightly, but only for a few days. However, in this case, temperature is the primary cause of these fluctuations, rather than damage from heavy rains or winds. In 2017 and 2019, average temperatures were higher than in 2018. In 2021, the average temperature was higher than in 2022.

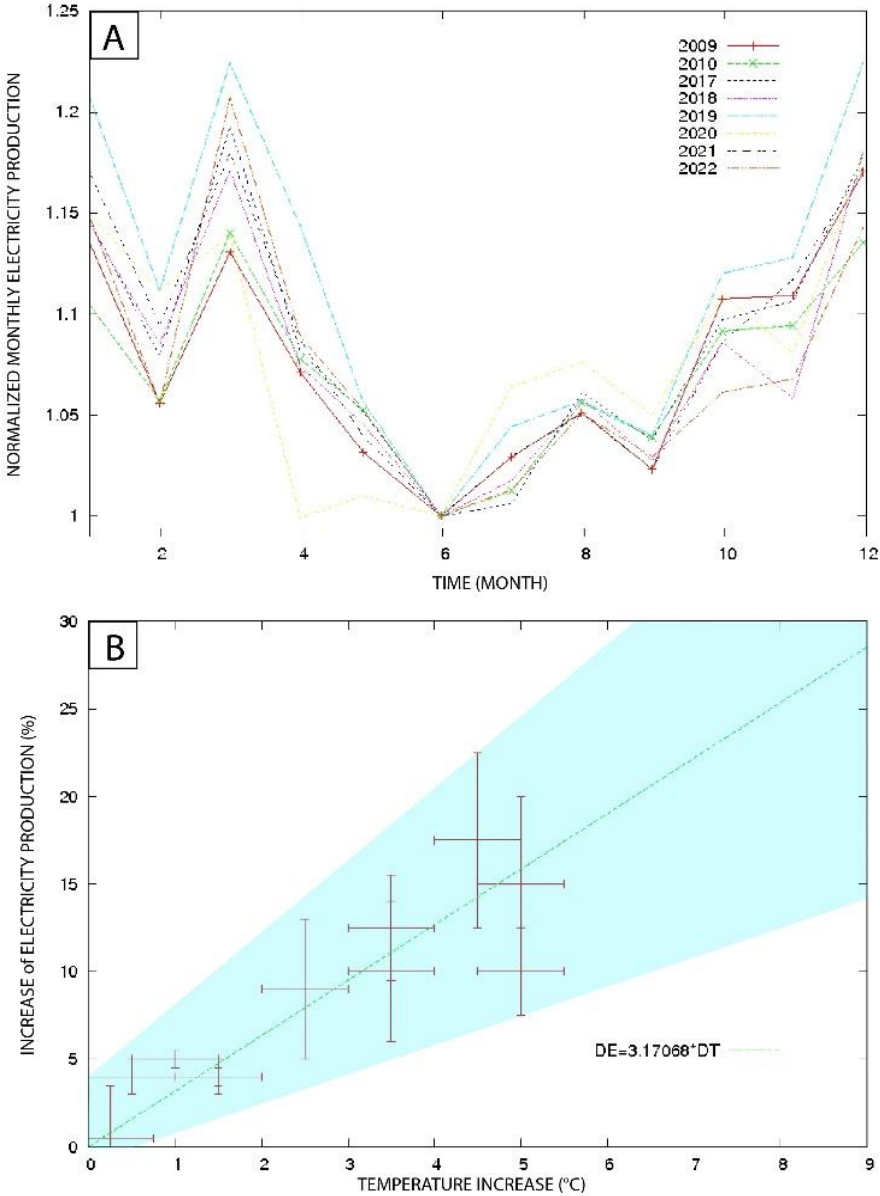

***Figure 10****: (A) Monthly electricity production in La Réunion normalized by the lower electricity production (June) for the years 2009, 2010, 2017, 2018, 2019, 2020, 2021, 2022. (B) Influence of temperature increase above 26°C on electricity production increase. Data source: Observatoire de l'énergie de La Réunion (REF). DE=3.17068\*DT (rms=2.53361, variance residual=6.419).*

The hottest year is 2019 (Table 3), and it also has the highest electricity consumption (Figure 10A). The seasonal increase in temperature causes an increase in energy consumption on La Réunion Island (Figure 10, Table 4). A 1°C increase in temperature above 26°C (i.e., from 31°C to 32°C) between December and March could increase electricity energy consumption by 3 to 6% (Figure 10B).

**Table 4**: *Influence of temperature variation of electricity production at La Réunion. Data in relation with figure 10. Data sources: Météo France (https://meteofrance.re/fr), Observatoire de l'Energie de la Réunion (https://oer.spl-horizonreunion.com/ ).*

| Month | % of Electricity Production E increase | Temperature T increase (in °C) | Uncertainty on E Variation (%) | Uncertainty on T |
|---|---|---|---|---|
| 1-january | 15 | 5 | 5 | 0.5 |
| 2-february | 10 | 5 | 2.5 | 0.5 |
| 3-march | 17.5 | 4.5 | 5 | 0.5 |
| 4-april | 12.5 | 3.5 | 3 | 0.5 |
| 5-may | 4 | 1.5 | 1 | 0.5 |
| 6-june | 0.5 | 0.25 | 3 | 0.5 |
| 7-july | 4 | 0.5 | 1 | 0.5 |
| 8-august | 5 | 1 | 0.5 | 0.5 |
| 9-september | 4 | 1.5 | 0.5 | 0.5 |
| 10-october | 9 | 2.5 | 4 | 0.5 |
| 11-november | 10 | 3.5 | 4 | 0.5 |
| 12-december | 17.5 | 4.5 | 5 | 0.5 |

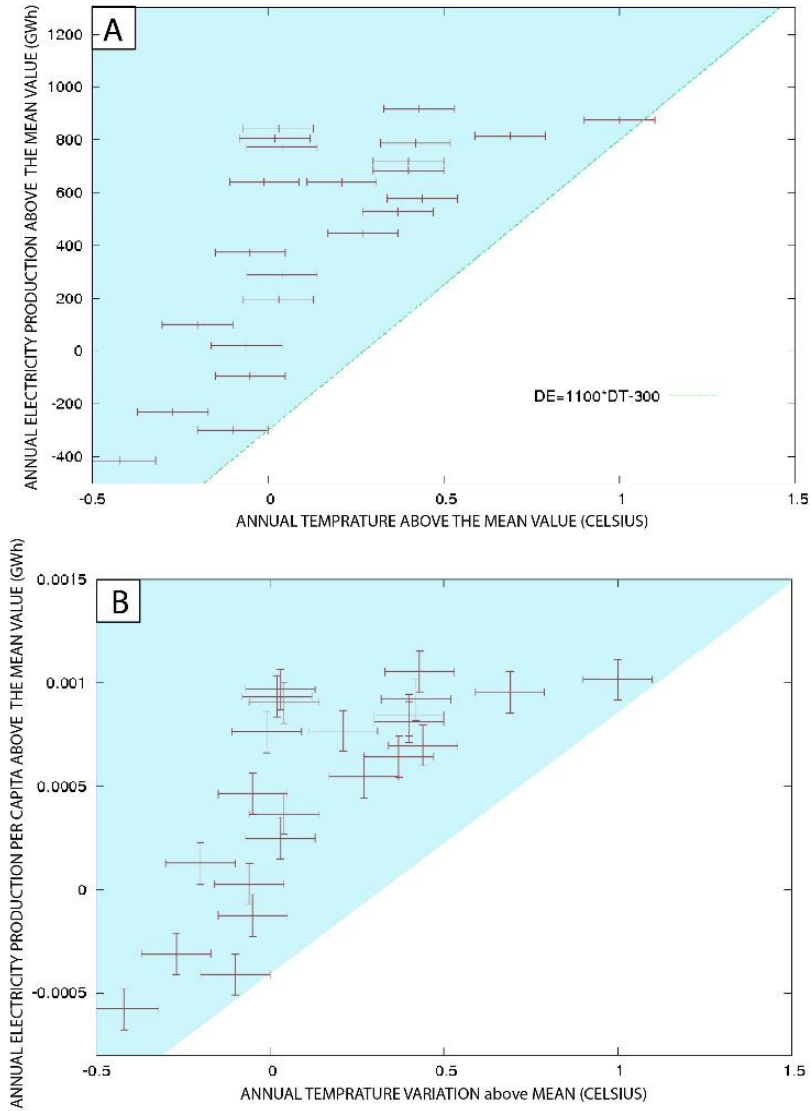

**FIGURE 11**: *Influence of annual temperature on annual electricity production in La Réunion island. (A) Annual Electricity production, above or below the mean value estimated between 1991 and 2020, as a function of the annual temperature variation, above the mean value estimated during the period 1991-2020, (B) Annual Electricity production per inhabitant, above the mean value estimated between 1991 and 2020, as a function of the annual temperature variation, above or below the mean value estimated during the period 1991-2020. The mean value of Electricity production from 1991 to 2020 (30 years) is 2170 GWh. Data source: the annual temperature above and below the mean temperature value estimated from 1991 to 2020 is estimated by Météo France (https://meteofrance.reu/fr). Data source: the annual electricity production from 1991 to 2022 is from Observatoire de l'énergie de La Réunion (https://oer.spl-horizonreunion.com/).*

Climate warming has caused an increase in average temperatures on La Réunion Island over the last fifty years (Météo France; *https://meteofrance.re/fr*). Electricity consumption per capita has increased in La Réunion Island over the last few decades, indicating that population growth is not the sole cause.

The increase in yearly energy production on La Réunion Island over the last few decades can be attributed in part to rising temperatures (Figure 11). As previously stated, another part of the increase in energy consumption was caused by the population increase, as well as by the techno-economic development and the wealth increase.

## 4. Interpretation and discussion

### 4.1 Climate driven energy consumption

The aim of this study is not to describe and discuss all interactions between the environment and society, but rather to focus on those involving energy production and consumption. It is well known that the production of fossil energy causes carbon dioxide emissions and climate warming [IPCC, 2013]. Variations in fossil energy consumption generate variations in the climate, as well as environment impact such as increased soil, water, and air pollution. The complexities of interactions between environment (soil, water, and air) and human behavior will not be discussed here, even if energy production is directly or indirectly involved in these pollutions.

The data presented in this study are related to energy production and consumption (Table 5). Consequently, the interpretation and discussion primarily focus on the role played by energy in these interactions. This study demonstrates that increasing temperatures causes an increase in energy consumption on both a monthly and yearly scale. The causal relationship between temperature increases and growth of energy consumption per capita in small tropical islands can be observed since the 1990s. This effect was not observed before 1990s in small tropical islands because: (i) it occurred during a period of strong economic, energy-demographic growth, (ii) hot events are more intense now, and (iii) the use of air conditioning primarily increased during the last two decades.

*Table 5*: *Relationship between natural or anthropic events and energy consumption.*

| Event | Energy consumption | Example |
|---|---|---|
| Cyclone | Decrease | Irma, 2017, Saint-Martin Ava and Dumazile, 2018, La Réunion |
| Temperature increase > 26°C | Increase | La Réunion |
| Migration out | Decrease | Saint-Martin, 2017 |
| Population increase | Increase | Small tropical islands |
| Wealth increase | Increase | Small tropical islands |
| Epidemic crisis | Decrease | Covid-19, Chikungunya |

The peaks in electrical consumption in small tropical islands are mainly due to the use of air conditioners. The peaks of electricity consumption in La Réunion Island are from 10h to 14h during austral summer, when the temperature increases, and from 17h to 19h during austral winter, at the end of the working day, when people return home and use household appliances. These observations are consistent with the interpretation of the role of temperature on electricity consumption. The increase in tourist arrivals during the dry season seems to have no significant influence on electricity consumption in La Réunion Island, even if tourism represents around 10% of the island's businesses and approximately 450,000 arrivals per year, mainly between June and November [IEDOM, 2014] (https://www.iedom.fr/IMG/pdf/ne293_eclairage_le_tourisme_a_la_reunion.pdf).

The consumption of small tropical islands is small compared to the rest of the world. Nevertheless, it could be used to broadly estimate the impact of increasing temperature in tropical areas on energy production. It is estimated that tropical area's population is 3.5 billion of inhabitants and represents 48% of the population of the Earth [Marcotullio et al., 2021]. A broad estimation of energy consumption of tropical territories is to consider that 700 more inhabitants consume 700 more energy (i.e. if 5 million of inhabitants consume 10.5 TWh, then 3.5 billion of inhabitants consume 7400 TWh). Assuming an energy consumption of 7400 TWh in tropical areas (and a yearly electricity consumption of 740 TWh, equivalent to 10% of the total energy consumption), it can be estimated that an increase of 1°C-2°C will cause an increase of 3% to 6% in electricity production, as observed in small tropical islands. This represents an increase of 22 to 44 TWh of electricity consumption per year for tropical areas when temperature increases of 1°C to 2°C.

The progressive increase of wealth in numerous countries, especially in tropical areas, may favor the increase of energy consumption. If there is no change in the trends demonstrated before, it is expected that energy consumption associated with the use of air conditioners will increase in tropical areas. Furthermore, the development of mobility and electric vehicles is expected to increase electricity consumption over the next few decades, even if in a non-equal way [Furszyfer Del Rio et al., 2023; Sadik Okah and Chidi Onuoha, 2024]. The fact that wealth correlates with an increase in electricity consumption in small tropical islands, does not imply that these rules will always apply. For example, it has been shown that GDP per capita has increased significantly in recent years, despite increases in electricity consumption was small.

Currently, the majority of electricity is produced using fossil resources. A 1°C or 2°C increase in mean temperature in tropical areas could significantly increase greenhouse gas emissions. In these regions, the feedback between global warming and electricity production is unsatisfactory and is expected to increase. Reducing fossil energy consumption, as well as energy consumption, is a necessary solution to reducing the anthropogenic impact of climate change.

In temperate areas, energy consumption increases significantly during the winter. Increasing temperatures may have different consequences on energy consumption depending on the context which must be investigated specifically. Nevertheless, climate change will have significant impact on energy production and consumption, if current behavior remain constant.

This study demonstrates that energy is a relevant indicator for monitoring societal changes, as well as to record the impacts of climate on society. More precisely, the monitoring of climate impact on society could be measured from energy consumption, because climate variations cause: (i) fluctuations in energy consumption due to the use of heating and air conditioning to regulate building temperatures, (ii) reductions in energy consumption due to reductions in economic, social and cultural activities, following the total or partial destruction of buildings and infrastructures by extreme hydro-climatic events, such as cyclones, (iii) energy variations when migration following crisis or disasters occurs, such as those that occurred on Saint-Martin Island.

### 4.2 Mitigation of negative effects

The present causal relationship between temperature increase and electrical consumption is not expected to be universal or eternal: first, it depends on the temperature that are observed, and this effect is certainly not observed for temperature <15°C; second, the increase of electricity consumption will be certainly lower if buildings are built with materials favoring isolation from sun heating and air conditioning electricity consumption will decrease with technological development; third, behavior may change and higher temperatures in buildings may be accepted in the future.

To reduce negative feedback, it is necessary to: (i) reduce carbonic energy production, (ii) boost economic activity without increasing air pollution and inequality. The Human Development Index (HDI) or the GDP/inhabitants could growth significantly even if energy production and consumption is not increasing, when strategies of energy consumption reduction are conducted [Radanne and Puiseux, 1989]. The lifestyles in wealthy territories generates an increased consumption of energy and environmental impacts, as suggested in previous studies [Pettifor et al, 2023; Gargani and Jouannic, 2015; Gargani, 2016b].

Causality must be considered carefully, not because it does not exist, but because changes in behaviors could modify the present observed trends that will not be identical in the future. In socio-economic cases, the causality is not identical than in physical cases. Social behaviors cannot be considered as physical laws that cannot be changed [Gargani, 2007]. The possibility to reduce negative retroaction between climate and energy must be explored.

Territories could choose to boost the Human Development Index rather than the GDP, thereby prioritizing social activities and care. The Human Development Index (HDI) has a positive correlation with an index that measures electricity consumption, such as the Night Light Development Index (NLDI) [Elvidge et al., 2012] suggesting that HDI increase may cause energy consumption increase, even if in a reduced way.

Inequality exists and are expending in the carbon footprint [Zheng et al., 2023]. Nevertheless, the trends that correlate HDI with energy consumption could be partly modified by changing behavior and improving technologies. There are various lifestyles that modify the carbon footprint and may decrease it [Pettifor et al., 2023]. Strategies and policies influence energy consumption.

The number of inhabitants influences significantly the energy consumption in the small tropical islands studied. Nevertheless, the possibility of reducing on energy consumption for a constant number of inhabitants should be considered as a possible objective: Energy consumption could be reduced, for example: (i) using appropriate materials to reduce thermal effects in buildings, (ii) considering alternative modes of transportation behavior (train, bicycle, walking, etc.), and (iii) considering efficient networks and electrical devices.

An increase or decrease in population does not result in better or worse energy use by the inhabitants on these islands: there is no trend between electricity consumption per inhabitants and the number of inhabitants. When the population grows, there appears to be no additional energy saving or energy waste. There is no obvious effect.

In the small tropical islands investigated, there is no relationship, at the present time, between the territory's GDP per capita and the development of renewable energies (Figure 12). The main renewable energies in small tropical islands are: (1) hydroelectricity, (2) geothermal energy, and

(3) solar energy. The percentage of renewable energy on small tropical islands depends on when renewable energy development began locally. Hydroelectricity was the first renewable energy developed primarily in the 1980s, when conditions were favorable (water and relief). Solar energy has only recently developed on these islands and will not account for a significant portion of total energy production by 2024.

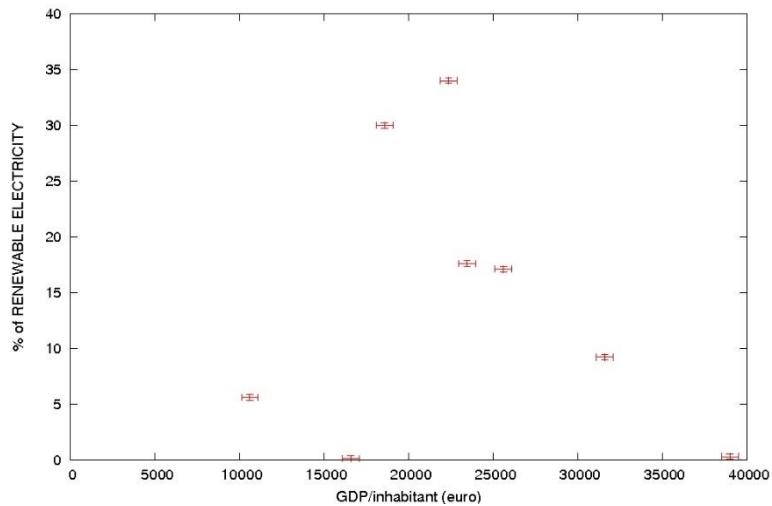

***FIGURE 12****: Influence of wealth on renewable electricity production.*

The fluctuations in energy production and consumption reflect interactions between the environment and society, but also geopolitical evolutions. The increase in renewable energy production, in the 1980s, is a reaction to the oil crisis of the 1970s [Radanne and Puiseux, 1989]. Recent conflicts also produced effects on oil and gas prices and energy consumption.

## Conclusion

Impact of climate on electricity consumption is observed in small tropical islands when temperature increase or extreme hydro-climatic events, such as cyclones, occur. The effects of climate on energy consumption in small tropical islands are sometime indirect, but not negligible. Climate change causes variations in energy consumption through the following effects: (i) the destruction of electricity-producing infrastructures, (ii) the destruction of electricity-consuming infrastructures, (iii) the reduction of electricity-consuming activities, (iv) the destruction or damage to networks, such as roads or telecommunications, (v) the increased use of air conditioning, and (vi) lifestyles change with air temperature and weather.

Socio-health events (covid-19), socio-demographic (increased population, migration), socio-technical (desire to reduce waste, energy savings, better building insolation) and socio-politic (economic crisis, new laws) all have a significant influence on energy consumption and may be linked to climate change.

## References

Akter S., 2023. Australia's Black Summer wildfires recovery: A difference-in-differences analysis using nightlights. Global Environmental Change, v.83, 102743.

Aramendia E., Brockway P.E., Taylor P.G., Norman J., 2023. Global energy consumption of the mineral mining

industry: Exploring the historical perspective and future pathways to 2060, Global Environmental Change, v.83, 102745.

ARER - Agence Régionale Energie Réunion, 2010. Rapport Etudes Consommations Energétiques des Ménages Réunionnais, pp.118.

Artelia, 2020. REX des consommations d'électricité des bâtiments tertiaires à La Réunion et en Guadeloupe, pp.91.

Bertram C., Luderer G., Creutzig F., Bauer N., Ueckerdt F., Malik A., Edenhofer O., 2021. COVID-19-induced low power demand and market forces starkly reduce $CO_2$ emissions. *Nat. Clim. Chang.* 11, 193–196. https://doi.org/10.1038/s41558-021-00987-x

Blaikie P.M., Cannon T., Davis I.,1994. *At Risk: Natural Hazards, People's Vulnerability, and Disasters*, Routledge, London; New York, 284 pp.

Blondeel M., Price J., Bradshaw M., Pye S., Dodds P., Kuzemko C., Bridge G., 2024. Global energy scenarios: A geopolitical reality check. Global Environmental Change, v.84, 102781.

Bolt J., Inklaar R., de Jong H., van Zanden J.L. Maddison Project Database, version 2018. Rebasing 'Maddison': new income comparisons and the shape of long-run economic development, 5, https://www.rug.nl/ggdc/historicaldevelopment/maddison/releases/maddison-project-database-2018

Bonneuil C., Fressoz J-C., 2013. *L'évènement Anthropocène : La Terre, l'histoire et nous*. Éditions du Seuil, Paris, 332 p.

Cangialosi J. P., Latto A. S., Berg R,. 2018. National Hurricane Center Tropical Cyclone Report. Hurricane Irma, AL112017.

Chardon J-P., Hartog T., 1995. Saint-Barthelemy : un choix et ses limites.

Cahiers d'outre-mer. N° 191 - 48e année, Juillet-septembre. Iles tropicales. p.261-276.

Chen Z.M., Chen G.Q., 2011. An overview of energy consumption of the globalized world economy. *Energy Policy*, v.39, pp.5920–5928.

Coumou D., Rahmstorf S., 2012. A decade of weather extremes. *Nature climate change*, v.2, n.7, 491.

Cutter S. L., 1996. "Vulnerability to environmental hazards", *Progress in Human Geography*, vol. 20, no. 4, pp. 529–539. doi: 10.1177/030913259602000407.

Cutter S.L., Boruff B.J., Shirley W.L., 2003. Social Vulnerability to Environmental Hazards. SOCIAL SCIENCE QUARTERLY, Volume 84, Number 2

Der Sarkissian, R., Dabaj, A., Diab, Y., Vuillet M., 2021. Evaluating the Implementation of the "Build- Back- Better" Concept for Critical Infrastructure Systems: Lessons from Saint- Martin's Island Following Hurricane Irma. *Sustainability*, 13, 3133.

Der Sarkissian R., Cariolet J-M., Diab Y., Vuillet M., 2022. Ivestigating the importance of critical infrastructures' interdependencies during recovery; lessons from Hurricane Irma in Saint-Martin's island. International Journal of Disaster Risk Reduction, v.67, 102675.

Deves M., Lacassin R., Pécout H., Robert G., 2022. Rick communication during seismo-volcanic crises: The example of Mayotte, France. Nat. Hazards Earth Syst. Sci., 22, 2001–2029.

Eco-concept Caraïbes, 2018. Bilan Assises de l'Environnement et de l'Energie : Saint-Barthélemy, pp.53.

EDF, 2015. Systèmes énergétiques insulaires : Martinique. Bilan prévisionnel de l'équilibre offre/demande d'électricité à Saint-Martin, pp.15.

EDF (Electricité De France), 2018a. Systèmes énergétiques insulaires de Saint-Barthélemy: bilan prévisionnel de l'équilibre offre/demande d'électricité, p. 1-7.

EDF (Electricité De France), 2018b. Systèmes énergétiques insulaires de Saint-Martin: bilan prévisionnel de l'équilibre offre/demande d'électricité, p. 1-7.

EDF, 2023a. Bilan prévisionnel de l'équilibre offre-demande d'électricité en Guadeloupe 2023-2028, pp. 14.

EDF, 2023b. Bilan prévisionnel de l'équilibre offre-demande d'électricité en Martinique 2023-2028, pp. 13.

EDF, 2023c. Bilan prévisionnel de l'équilibre offre-demande d'électricité à Saint-Barthélemy 2023-2028, pp. 17.

EDF, 2023d. Bilan prévisionnel de l'équilibre offre-demande d'électricité à Saint-Martin 2023-2028, pp. 17.

Elvidge C.D., Baugh K.E., Anderson S.J., Sutton P.C., Ghosh T., 2012. The Night Light Development Index (NLDI): a spatially explicit measure of human development from satellite data. Social Geography, 7, 23-35.

Fressoz J.-B., 2024. Sans transition Une nouvelle histoire de l'énergie. Seuil.

Furszyfer Del Rio J., Furszyfer Del Rio D.D., Sovacool B.K., Griffiths S., 2023. The demographics of energy and mobility poverty: Assessing equity and justice in Ireland, Mexico, and the United Arab Emirates, Global Environmental Change, v.81, 102703.

García S., Parejo A., Personal E., Guerrero J.I., Biscarri F., León C., 2021. A retrospective analysis of the impact of the COVID-19 restrictions on energy consumption at a disaggregated level, Applied Energy, v.287, 116547.

Gargani J., De la convivialité entre scientifiques. *La revue du MAUSS, n.29, p.127-156, 2007.*

Gargani J., G. Jouannic, 2015. Les liens entre Société, Nature et Technique durant les derniers 200 ans : analyse dans deux vallées françaises. *VertigO, V. 15, n.3, 2015.*

Gargani J., 2016a. Crises environnementales et crises socio-économiques. L'Harmatan, Paris.

Gargani J., 2016b. De la conception de la nature à la conception de la technique. *Sociétés, v.131, p.19-39.*

Gargani J, 2020. Modelling the mobility and dynamics of a large Tahitian landslide using runout distance. *Geomorphology, v.370, 107354.*

Gargani J., 2022a. Inequality growth and recovery monitoring after disaster using indicators based on energy production: Case study on Hurricane Irma at the Caribbean in 2017. *International Journal of Disaster Risk Reduction, v.79, 103166.*

Gargani J., 2022b. Impact of major hurricanes on electricity energy production. *Intern. Jour. of Disaster Risk Reduction, v.67, 102643.*

Gargani J. 2022c. Isostatic Adjustment, Vertical Motion Rate Variation and Potential Detection of Past Abrupt Mass Unloading. *Geosciences, 12(8), 302.*

Gargani J., 2023. Influence of Relative Sea-Level Rise, Meteoric Water Infiltration and Rock Weathering on Giant Volcanic Landslides. *Geosciences, 13(4), 113.*

Gargani J., 2024. Relative sea level and coastal vertical movements in relation to volcano-tectonic processes, Geohazards, 5, 329-349.

Halbrügge S., Schott P., Weibelzahl M., Buhl H.U., Fridgen G., Schöpf M., 2021. How did the German and other European

electricity systems react to the COVID-19 pandemic? *Applied Energy*, v.285, 116370.

Hachimi Alaoui M., Lemercier E., Palomares E., 2013. Reconfigurations ethniques à Mayotte. Hommes & Migration, 1304. Pp.9.

Hall C.AS, Klitgaard K.A., 2011. *Energy and the wealth of nations: understanding the biophysical economy*. Springer Science & Business Media.

Heady, D. D., Hodge, A., 2009. The effect of population growth on economic growth: A meta-regression analysis of the macroeconomic literature. Population and Development Review, 35, 221-248.

Howell, J., Elliott, J. R. (2019). Damages done: The longitudinal impacts of natural hazards on wealth inequality in the United States. *Social Problems*, *66*(3), 448-467.

Huang R., Lixin Tian L., 2021. CO2 emissions inequality through the lens of developing countries. *Applied Energy*, v.281, 116043.

IEDOM, (Institut d'Emission des Départements d'Outre-Mer), 2010. Rapport annuel Polynésie française, pp.231.

IEDOM, (Institut d'Emission des Départements d'Outre-Mer), 2012. Saint-Barthélemy annual report 2011, pp.96.

IEDOM, 2014. Le tourisme à La Réunion: une activité qui peine à décoller. Note expresse, n.293, 6 p., https://www.iedom.fr/IMG/pdf/ne293_eclairage_le_tourisme_a_la_reunion.pdf

IEDOM, (Institut d'Emission des Départements d'Outre-Mer), 2015. Rapport annuel Mayotte, pp.205.

IEDOM (Institut d'Emission des Départements d'Outre-Mer), 2020a. Saint-Barthelemy annual report 2019, pp.94.

IEDOM (Institut d'Emission des Départements d'Outre-Mer), 2020b. Saint-Martin annual report 2019, pp.104.

IEDOM (Institut d'Emission des Départements d'Outre-Mer), 2021a. Saint-Barthelemy annual report 2020, pp.104.

IEDOM (Institut d'Emission des Départements d'Outre-Mer), 2021b. Saint-Martin annual report 2020, pp.116.

IEDOM, (Institut d'Emission des Départements d'Outre-Mer), 2023a. Rapport annuel Guadeloupe, pp.217.

IEDOM, (Institut d'Emission des Départements d'Outre-Mer), 2023b. Rapport annuel Martinique, pp.232.

IEDOM, (Institut d'Emission des Départements d'Outre-Mer), 2023c. Rapport annuel Mayotte, pp.196.

IEDOM, (Institut d'Emission des Départements d'Outre-Mer), 2023d. Rapport annuel Polynésie française, pp.200.

IEDOM (Institut d'Emission des Départements d'Outre-Mer), 2023e. Saint-Barthelemy annual report 2022, pp.126.

IEDOM (Institut d'Emission des Départements d'Outre-Mer), 2023f. Saint-Martin annual report 2022, pp.144.

IEOM, 2013. Rapport annuel Nouvelle-Calédonie, pp.208.

IEOM, 2016. Rapport annuel Nouvelle-Calédonie, pp.184.

IEOM, 2019. L'économie verte en Nouvelle-Calédonie: un essor mesuré mais prometteur; n.2171, pp.10.

IEOM, 2020. Rapport annuel Nouvelle-Calédonie, pp.216.

IEOM, 2023a. Rapport annuel économique de La Réunion, 2022, pp.215.

IEOM, 2023b. Rapport annuel économique de la Nouvelle-Calédonie, 2022, pp.224.

INSEE, 2024. Données locales. Dossier complet, Commune de Saint-Martin (97127). pp. 6.

IPBES, 2019: Summary for policymakers of the global assessment report on biodiversity and ecosystem services of the Intergovernmental Science-Policy Platform on Biodiversity and Ecosystem Services. S. Díaz, J. Settele, E. S. Brondízio E.S., H. T. Ngo, M. Guèze, J., Agard, A. Arneth, P. Balvanera, K. A. Brauman, S. H. M. Butchart, K. M. A. Chan, L. A. Garibaldi, K. Ichii, J. Liu, S. M. Subramanian, G. F. Midgley, P. Miloslavich, Z. Molnár, D. Obura, A. Pfaff, S. Polasky, A. Purvis, J. Razzaque, B. Reyers, R. Roy Chowdhury, Y. J. Shin, I. J. Visseren-Hamakers, K. J. Willis, and C. N. Zayas (eds.). IPBES secretariat, Bonn, Germany.

IPCC, 2013. Climate change 2013: The physical science basis. Contribution of Working Group I to the Fifth assessment report of the Intergovernmental Panel on Climate Change [Stocker, T.F., Qin, D, Plattner, G.- K., Tignor, M., Aller, S.K., Boschung, J., Nauels, A., Xia, Y., Bex, V. et Midgley P.M. (eds.)]. Cambridge University Press, Cambridge, United Kingdom and New York, NY, USA, 1535p.

Jiang P., Fan Y.V., Klemeš J.J., 2021. Impacts of COVID-19 on energy demand and consumption: Challenges, lessons and emerging opportunities. *Applied Energy*, Volume 285, 116441.

Jones R.V., A. Fuertes, K.J. Lomas, 2015. The socio-economic, dwelling and appliance related factors affecting electricity consumption in domestic buildings**.** *Renewable Sustainable Energy Rev.*, 43, pp. 901-917.

Jouannic G., A. Ameline, K. Pasquon, O. Navarro, C. Tran Duc Minh, A. Halim Boudoukha, M-A. Corbillé, D. Crozier, G. Fleury-Bahi, J. Gargani, P. Guéro, 2020. Recovery of the Island of Saint Martin after Hurricane Irma: An Interdisciplinary Perspective. *Sustainability,* 12, 8585, doi:10.3390/su12208585.

Khan S.A.R., K. Zaman, Y. Zhang, 2016. The relationship between energy-resource depletion, climate change, health resources and the environmental Kuznets curve: Evidence from the panel of selected developed countries. *Renewable and Sustainable Energy Reviews*, v.62, p.468-477.

Kishore N., Marque´s D., Mahmud A., Kiang MV., Rodriguez I., Fuller A. et al., 2018. Mortality in Puerto Rico after Hurricane Maria. *New England journal of medicine*, Jul 12, 379, 2, 162–170.

Latouche S., 2004. *La Megamachine : Raison technoscientifique, raison économique et mythe du progrès*. Éditions La Découverte, Paris, 2002 p.

Li M., Allinson D., He M., 2018. Seasonal variation in household electricity demand: A comparison of monitored and synthetic daily load profiles, *Energy and Buildings*, v.179, 292-300.

Marcotullio P.J., Keßler C., Quintero Gonzalez R., Schmeltz M., 2021. Urban Growth and Heat in Tropical Climates. Frontiers in Ecology and Evolution, v.9. DOI=10.3389/fevo.2021.616626

Meyer T., 2021. Quelle transition énergétique en Polynésie française ? Géoconfluences, pp.15. https://geoconfluences.ens-lyon.fr/informations-scientifiques/dossiers-regionaux/la-france-desterritoires-en-mutation/articles-scientifiques/polynesie-transition-energetique

Navon A., Machlev R., Carmon D., Onile A.E., Belikov J., Levron Y., 2021. Effects of the COVID-19 Pandemic on Energy Systems and Electric Power Grids—A Review of the Challenges Ahead. *Energies*, 14, 1056.

Observatoire de l'énergie de La Réunion, 2011. Bilan énergétique de l'île de La Réunion, pp.60.

Observatoire de l'énergie de La Réunion, 2013a. Bilan énergétique 2012 de l'île de La Réunion, pp.59.

Observatoire de l'énergie de La Réunion, 2013b. Etude de la petite climatisation, période 2011-2012. pp.37.

Observatoire Polynésien de l'énergie, 2017. Bilan énergétique de la Polynésie française, pp.83.

Observatoire territorial de la transition ecologique et energetique, 2022. Bilan énergétique 2021 de la Martinique, pp.16.

Pasquon K., J. Gargani, G. Jouannic, 2019. Interaction nature/société de 1947 à 2017 : processus, caractéristiques et vulnérabilité à Saint-Martin. Proceedings of the Vulnérabilité et Résilience Conférence, November, Université Versailles-Saint-Quentin, France, p.1-14.

Pasquon K., Jouannic G., Gargani J., Tran Duc Minh C., Crozier D., 2022a. Urban evolution and exposure to cyclonic hazard in Saint-Martin between 1954 and 2017. Vertigo, v.22, n.1.

Pasquon K., Gargani J., Jouannic G. 2022b. Vulnerability to marine inundation in Caribbean islands of Saint-Martin and Saint-Barthélemy. International Journal of Disaster Risk Reduction, v.78, 103139.

Petrovics D., Huitema D., Giezen M., Vis B., 2024. Scaling mechanisms of energy communities: A comparison of 28 initiatives. Global Environmental Change, v.84, 102780.

Pettifor H., Agnew M., Wilson C., 2023. A framework for measuring and modelling low-carbon lifestyles. Global Environmental Change, v.82, 102739.

Pichler, A., E. Striessnig. 2013. Differential vulnerability to hurricanes in Cuba, Haiti, and the Dominican Republic: the contribution of education. Ecology and Society, vol. 18; n.3, p.31.

Piketty T., 2013. Le capital au 21e siècle. Edition du Seuil, Paris, pp.970.

Préfet de la Réunion, 2019. Programmation pluriannuelle de l'énergie (PPE), pp.11.

Préfet de Mayotte, 2016. Programmation pluriannuelle de l'énergie de Mayotte, pp.152.

Prina M.G., Groppi D., Nastasi B., Garcia D.A., 2021. Bottom-up energy system modelsapplied to sustainable islands. Renewable and Sustainable Energy Reviews. v.152, 111625.

Radanne P., Puiseux L., 1989. L'énergie dans l'economie. Alternatives économiques, Ed. Syros, pp.176.

Román MO., Stokes EC., Shrestha R., Wang Z., Schultz L., Carlo EAS. et al., 2019. Satellite-based assessment of electricity restoration efforts in Puerto Rico after Hurricane Maria. PLoS ONE, 14, 6, e0218883.

Rubin O, Rossing T., 2012. National and Local Vulnerability to Climate-Related Disasters in Latin America: The Role of Social Asset-Based Adaptation. Bulletin of Latin American Research. 31, 1, p.19–35. PMID: 22216472

Sadiq Okoh A., Chidi Onuoha M., 2024. Immediate and future challenges of using electric vehicles for promoting energy efficiency in Africa's clean energy transition. Global Environmental Change, v.84, 102789.

Salaün M., Trépied B., 2024. Nouvelle-Calédonie, une histoire de la colère. Le Monde diplomatique, juillet p.20-21.

Sovacool B.K., Hess D.J., Cantoni R., Lee D., Brisbois M.C., Walnum H.J., Dale R.F., Rygg B.J., Korsnes M., Goswami A., Kedia S., Goel S., 2022. Conflicted transitions: Exploring the actors, tactics, and outcomes of social opposition against energy infrastructure, Global

Environmental Change, Volume 73, 2022, 102473.

Stiglitz J., Fitoussi J-P., Durand M., 2018. Beyond GDP: Measuring What Counts for Economic and Social Performance, Sciences Po publications info:hdl:2441/4vsqk7docb9, Sciences Po.

Syndicat des énergies renouvelables, 2018. Autonomie énergétiqu en 2030 pour les outre-mer et la Corse, pp.12.

Syvitski, J., Waters, C.N., Day, J. *et al.* 2020. Extraordinary human energy consumption and resultant geological impacts beginning around 1950 CE initiated the proposed Anthropocene Epoch. *Commun Earth Environ,* v.1, n.32. https://doi.org/10.1038/s43247-020-00029-y

Tsimanda F.I., 2023. Migrer pour un bidonville. La vulnérabilité socio-économique des migrants comoriens à Mayotte. Géoconfluences. https://geoconfluences.ens-lyon.fr/informations-scientifiques/dossiersthematiques/inegalites/articles/migrants-comores-mayotte.

Van der Borght R., Pallares-Barbera M., 2024. Greening to shield: The impacts of extreme rainfall on economic activity in Latin American cities, Global Environmental Change, v.87, 102857.

Zheng H., Wood R., Moran D., Feng K., Tisserant A., Jiang M., Hertwich E.G., 2023. Rising carbon inequality and its driving factors from 2005 to 2015, Global Environmental Change, v.82, 102704.